\documentclass[aps,twocolumn,showpacs,floatfix]{revtex4}
\usepackage{bm}
\usepackage{graphicx}
\usepackage{epsfig}
\usepackage{amsmath} 
\usepackage{color}
\usepackage{ulem}

\begin{document}

\title{Split Dirac cones in HgTe/CdTe quantum wells due to symmetry-enforced level anticrossing at interfaces}

\author{S.\,A.\,Tarasenko}
\author{M.\,V.\,Durnev}
\author{M.\,O.\,Nestoklon}
\author{E.\,L.\,Ivchenko}

\affiliation{A.\,F.\,Ioffe Physical-Technical Institute, 194021 St.\,Petersburg, Russia}

\author{Jun-Wei Luo} 
\affiliation{State Key Laboratory of Superlattices and Microstructures, Institute of Semiconductors, Chinese Academy of Sciences, Beijing 100083, China,
Synergetic Innovation Center of Quantum Information and Quantum Physics, University of Science and Technology of China, Hefei, Anhui 230026, China}

\author{ Alex Zunger}
\affiliation{University of Colorado, Boulder, Colorado 80309, USA}

\begin{abstract}
We describe the fine structure of Dirac states in HgTe/CdHgTe quantum wells of critical and close-to-critical thickness and demonstrate 
the formation of an anticrossing gap between the tips of the Dirac cones
driven by interface inversion asymmetry. By combining symmetry analysis, atomistic calculations, and $\bm k$$\cdot$$\bm p$ theory with interface terms, we obtain a quantitative description of the energy spectrum and extract the interface mixing coefficient. The zero-magnetic-field splitting of Dirac cones can be experimentally revealed in studying magnetotransport phenomena, cyclotron resonance, Raman scattering, or THz radiation absorption.
\end{abstract}





\maketitle 

The study of systems with gapless and linear-dispersion states constituting the Dirac cone is central to the physics of topological insulators (TIs)~\cite{Hasan2010,Qi2011}. Such states are formed in the primary insulating band gap of the bulk material and studied at the surface via techniques such as angular resolved photoemission~\cite{Xia2009,Tanaka2012}. While some TI compounds such as Bi$_2$Se$_3$ and Bi$_2$Te$_3$ show an insulating gap in their three-dimensional (3D) bulk band structure and so the Dirac cones can be studied at their 2D surface without modifying the material, a Mercury Telluride crystal does not have a 3D bulk band gap because the Fermi level resides within the four-fold degenerate $\Gamma_8$ band~\cite{Dyakonov1981}. However, the topological insulation is realized by straining HgTe which opens a gap within the otherwise fourfold degenerate $\Gamma_8$ states~\cite{Bruene2011,Kozlov2014} or growing the material in a HgTe/CdTe quantum well (QW) geometry~\cite{Bernevig2006,Konig2007,Luo2010}. In the latter case, CdTe (or Cd$_{1-x}$Hg$_x$Te) barriers create quantum confinement within HgTe with the normal or inverted band structure depending on the well thickness, so that HgTe/CdTe QWs belong to the class of normal or topological insulators.

\begin{figure}[t]
  \includegraphics[width=0.75\columnwidth]{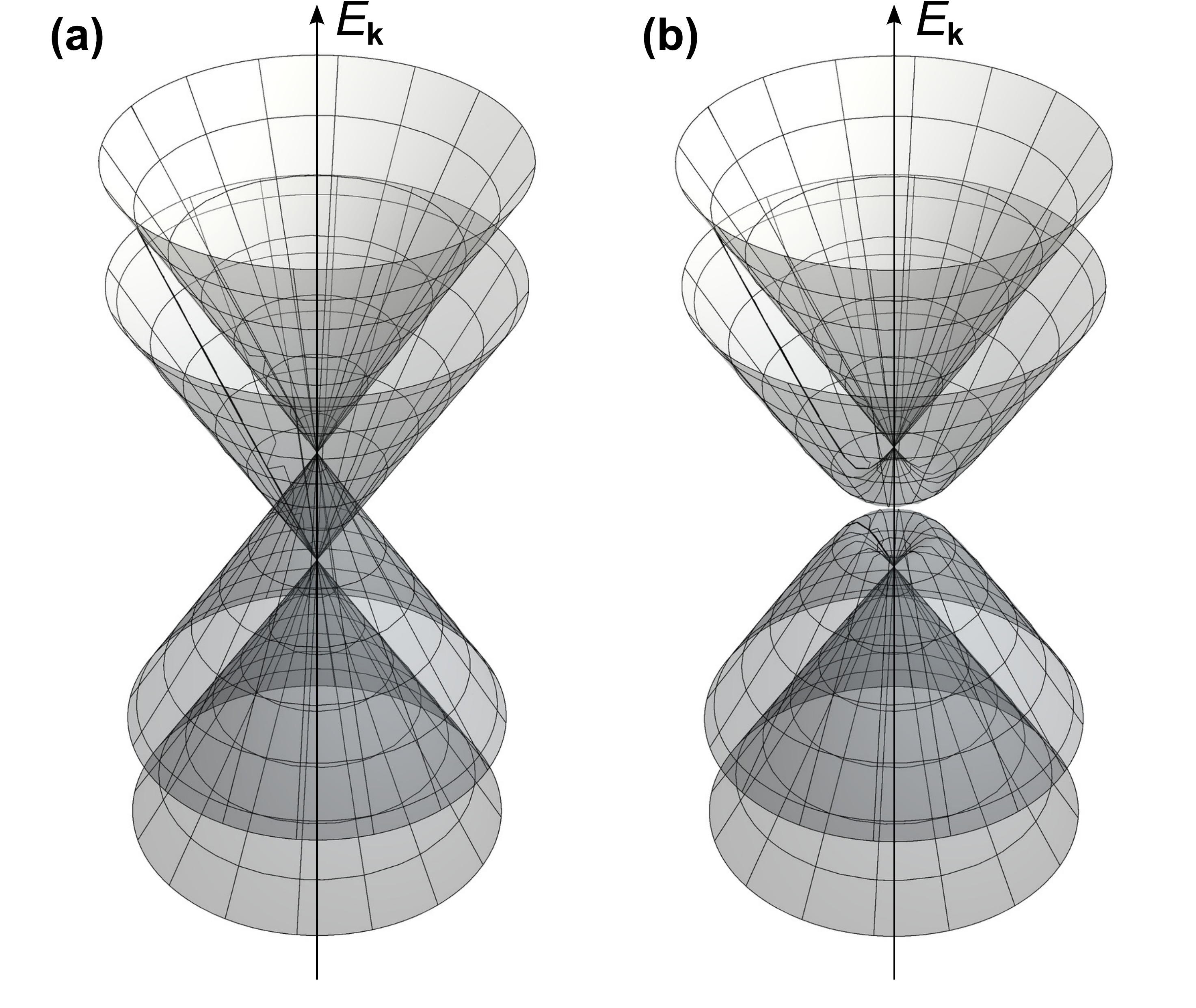}
\caption{Energy spectra of HgTe/CdHgTe QWs of (a) the critical and (b) close-to-critical thicknesses. The spectra are plotted after Eq.~\eqref{dispersion} for (a) $\delta=0$ and (b) $\delta=\gamma/2$.}
\label{fig1}
\vspace{-0.5cm}
\end{figure} 

In HgTe/CdTe  QWs of critical thickness, the heavy-hole subband $HH1$ switches order with the  electron subband $E1$~\cite{Gerchikov1989}, the band gap vanishes and elementary excitations behave as massless Dirac 2D fermions~\cite{Buttner2011,Olbrich2013}. The early theoretical description~\cite{Bernevig2006} used a model in which the symmetry of the HgTe/CdTe QW was implicitly assumed to contain an inversion center. Consequently, the point-group representations of the $E1$ and $HH1$ subbands are different, so these states do not mix with each other and are allowed to cross. In reality, the host zinc-blende structures lack the inversion center and the QW structure has a further reduced symmetry compared to the bulk materials even if the interfaces are lattice-matched and defect-free, and the well is symmetric. Indeed, each of the two (001) interfaces possesses a C$_{2v}$ symmetry and taken together the symmetric well has the D$_{2d}$ symmetry. It is known from analogous GaAs/AlAs QW structures that, in the D$_{2d}$ group, the $E1$ and $HH1$ states transform according to the same spinor representations~\cite{Ivchenko1996,Krebs1998,Magri2000}. Therefore, the coupling matrix element between $E1$ and $HH1$ is nonzero, and these subbands must anticross rather than cross at zero in-plane wave vector $\bm{k}$. 
As pointed out in Refs.~\cite{Dai2008,Konig2008,Winkler20012,Weithofer2013} in the case of HgTe/CdTe QWs a reduction in symmetry leads to the splitting of states at $\bm{k} = 0$. However, a theory identifying the source of the splitting (bulk vs interface) and predicting its magnitude has been lacking. Experimental data on weak localization and Shubnikov$-$de Haas oscillations also indicate strong spin-orbit splitting of the states~\cite{Minkov2014}. Here, we present a microscopic theory of the Dirac states in HgTe/CdTe QWs that predicts a very large ($\sim15$ meV) anticrossing gap between the tips of the Dirac cones in QWs of the critical thickness. We find that this splitting is predominantly due to the interfacial $E1$-$HH1$ repulsion mandated by the physical $D_{2d}$ symmetry which is missed by naive continuum-medium considerations but seen when theory acquires atomic resolution. Using this picture we further provide a detailed analysis of the fine structure of Dirac states, which is a key to understanding transport phenomena. As the main result, Fig.~\ref{fig1} shows the energy spectra in QWs of (a) the critical thickness, $d = d_c$, and (b) close-to-critical thickness, $d \neq d_c$. Even in symmetric structures of the critical thickness, the Dirac states are split at the zone center and the spectrum consists of two cones shifted vertically with respect to each other. For $d \neq d_c$, the spectrum becomes more complex: It has a gap and extrema on a circle in the momentum space. We analyze the nature of anticrossing, obtain a quantitative description of the energy spectrum  and discuss the consequences of the splitting on transport and optical properties of QWs. 

Mechanisms leading to the state splitting in QWs with symmetric confinement potential originate fundamentally from bulk inversion asymmetry of the host crystal and interface inversion asymmetry. The relative importance of these contributions cannot be deduced from model Hamiltonian consideration. Atomistic descriptions, on the other hand, capture accurately the relevant asymmetry via the specification of atomic types and positions. 

\begin{figure}
 \includegraphics[width=\columnwidth]{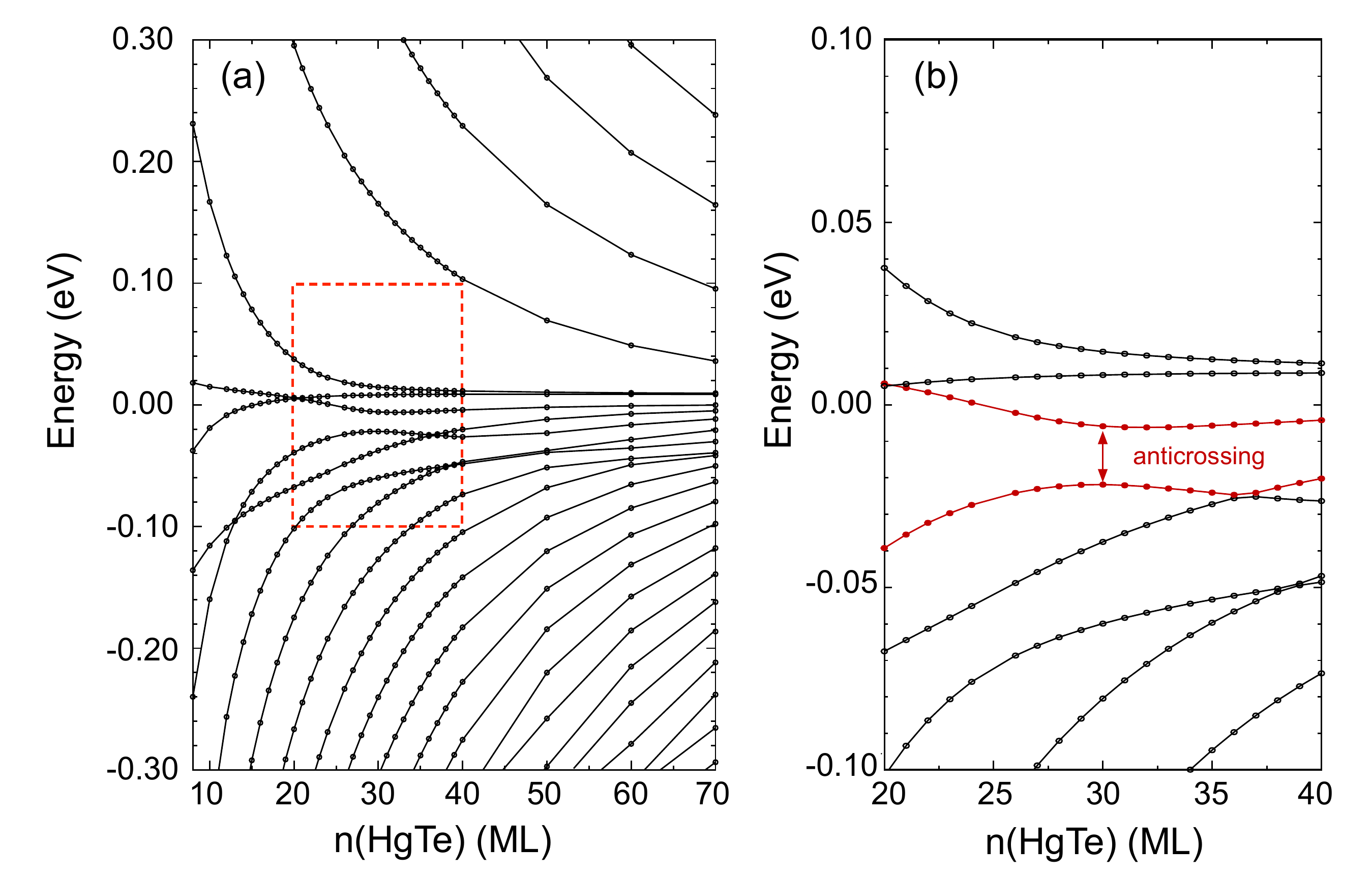}
\caption{(a) Arrangement of energy subbands in (001) (HgTe)$_n$/(CdTe)$_{40}$ QW structures as a function of the number of HgTe monolayers obtained by the pseudopotential method. (b) Zoom in of the anticrossing area. 
}
\label{fig2}
\end{figure}

\begin{figure}
 \includegraphics[width=\columnwidth]{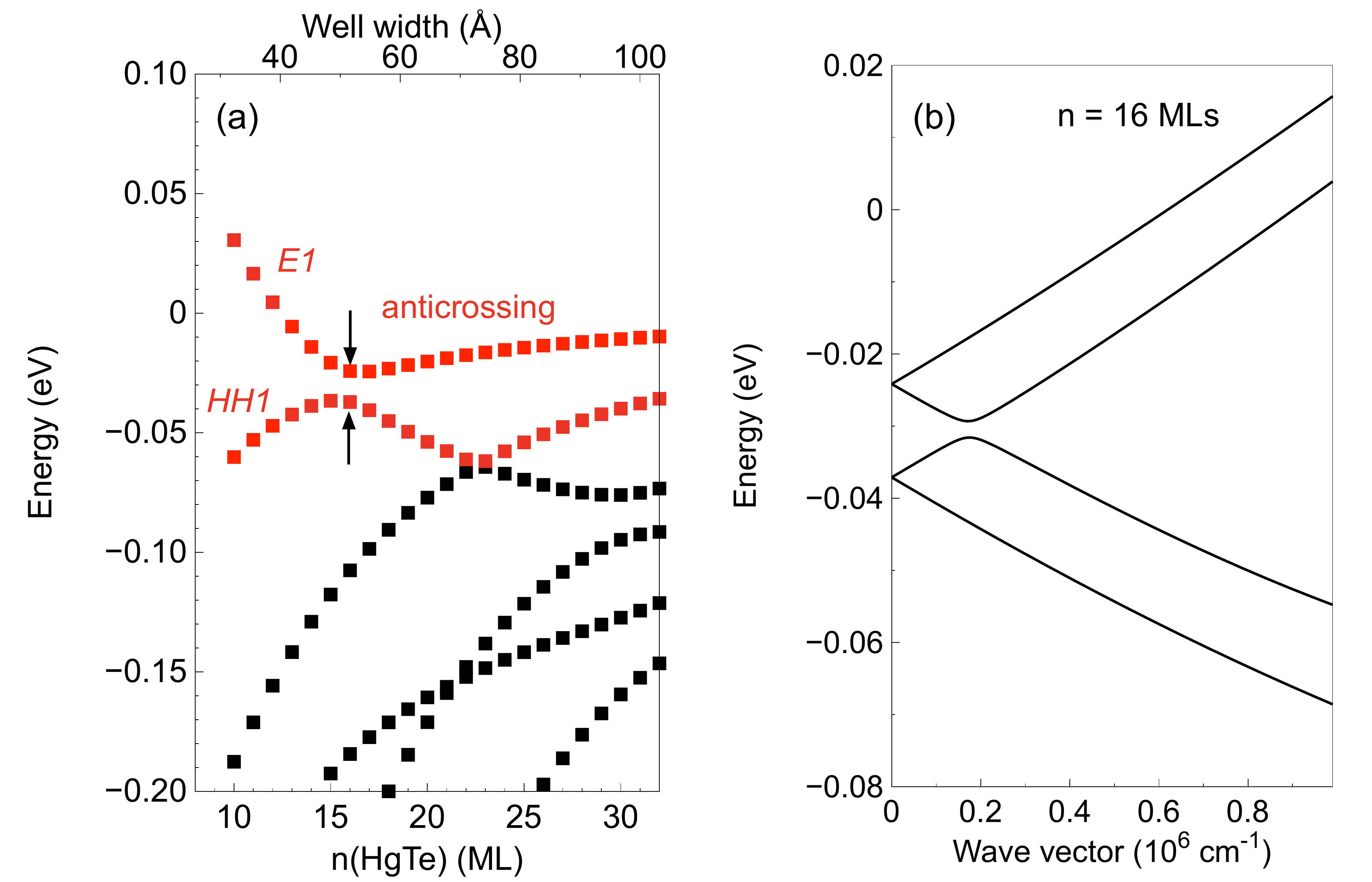}
\caption{(a) Arrangement of energy subbands in (001) (HgTe)$_n$/CdTe QW structures as a function the number of HgTe monolayers obtained by the $sp^3$ tight-binding method. (b) Electron dispersion $E(\bm k)$ in the 16-ML-wide QW.  
}
\label{fig3}
\end{figure}

Figures~\ref{fig2} and \ref{fig3} show the results of atomistic calculations of the energy spectrum of HgTe/CdTe QWs obtained in the screened plane-wave pseudopotential and tight-binding theories, respectively. The pseudopotential method is described in Ref.~\cite{Luo2010}; details of the tight-binding calculations are given in Supplementary Material. The subband arrangement as a function of the QW width at $\bm k = 0$ is presented in Figs.~\ref{fig2}a, \ref{fig2}b and \ref{fig3}a. Both the pseudopotential and tight-binding calculations yield a wide anticrossing gap between the electron-like $E1$
and heavy-hole $HH1$ subbands at the $\Gamma$ point. The two approaches give different values of the critical
QW width because of the sensitivity of the subband structure to the model. Indeed, the pseudoptential calculation reveals for the QW structures interface-localized bands located energetically between $E1$ and the $HH1$ state, see Ref.~\cite{Luo2010}.  
Nevertheless, all calculations that acknowledge the atomistic symmetry -- as opposed to a continuum view -- do give large anticrossing at the critical QW thickness. They both predict the gap of about 15 meV far exceeding the estimate of a few meV due solely to the bulk inversion asymmetry~\cite{Konig2008, Winkler20012} and unambiguously indicating that the subband
mixing is dominated by the interface contribution, in a crutial difference with the naive $\bm k$$\cdot$$\bm p$ model. Moreover, despite the existence of an additional interface bands in the pseudopotential calculation, the both models predict the same dispersion of the Dirac states formed from the $E1$ and $HH1$ subbands near the anticrossing point. 

Figure~\ref{fig3}b demonstrates the energy dispersion $E({\bm k})$ of the coupled $E1$ and $HH1$ states calculated by the tight-binding method for the well of 16 monolayers (ML) in width which is close to the critical thickness. The spectrum consists of four almost linearly dispersed branches which are split at $k=0$. The slope of the branches yields the velocity $6.1 \times 10^7$~cm/s which is close to the electron velocity $7.2 \times 10^7$~cm/s~\cite{Olbrich2013} and $6.4 \times 10^7$~cm/s~\cite{ Ludwig2014} determined by cyclotron resonance in HgTe/HgCdTe QWs of critical thickness.
In Fig.~\ref{fig3}b, the two middle-energy branches anticross at the finite wave vector $k_0 \approx 0.17 \times 10^6$~cm$^{-1}$ with the energy gap being highly sensitive to the deviation of QW width $d$ from $d_c$. In QWs with $d \approx d_c$ (the exact condition $d=d_c$ cannot be fulfilled in structures with an integer number of monolayers), this gap is far smaller than the splitting at $k = 0$. 

Inspired by the atomistic results we now present an effective model Hamiltinian -- the atomically inspired $\bm k$$\cdot$$\bm p$ model (AIKP) -- which takes into account the correct $D_{2d}$ symmetry of the quantum well. Limiting the AIKP model to four basis Bloch functions, which form the Dirac states, it can be solved analytically.
The basis functions have the form~\cite{Bernevig2006} 
\begin{eqnarray} \label{Bloch}
&&| E1,+1/2 \rangle = f_1(z) |\Gamma_6,+1/2 \rangle + f_4(z) |\Gamma_8,+1/2 \rangle \:, \nonumber \\
&&| HH1,+3/2 \rangle = f_3(z) |\Gamma_8,+3/2 \rangle \:, \nonumber \\
&&| E1,-1/2 \rangle = f_1(z) |\Gamma_6,-1/2 \rangle + f_4(z) |\Gamma_8,-1/2 \rangle \:, \nonumber \\
&&| HH1,-3/2 \rangle = f_3(z) |\Gamma_8,-3/2 \rangle \:, 
\end{eqnarray}
where $f_1(z)$, $f_3(z)$, and $f_4(z)$ are the envelope functions, $z$ is the growth direction, $|\Gamma_6,\pm 1/2 \rangle$, $|\Gamma_8,\pm 1/2 \rangle$, and $|\Gamma_8,\pm 3/2 \rangle$ are the basis functions of the $\Gamma_6$ and $\Gamma_8$ states at the $\Gamma$ point of the Brillouin zone, respectively. The $4\times 4$ effective Hamiltonian describing the coupling at a finite in-plane wave vector $\bm k$ can be constructed by using the theory of group representations. Taking into account that, in the 
$D_{2d}$ point group, the states $|E1,\pm 1/2 \rangle$ and $|HH1, \mp 3/2 \rangle$ transform according to the spinor representation $\Gamma_6$ while the wave vector components $k_x,k_y$ belong to the irreducible representation $\Gamma_5$, one derives the effective Hamiltonian to first order in the wave vector as
\begin{equation}\label{H_eff}
H = \left( 
\begin{array}{cccc}
\delta & {\rm i} A k_+ & 0 & {\rm i} \gamma \\
-{\rm i} A k_- & -\delta & {\rm i} \gamma & 0 \\
0 & -i\gamma & \delta & - {\rm i} A k_- \\
-{\rm i}\gamma & 0 & {\rm i} A k_+ & -\delta
\end{array}
\right)\,\, ,
\end{equation}
where $A$, $\delta$, and $\gamma$ are linearly independent parameters, $k_{\pm} = k_x \pm {\rm i} k_y$, $x \parallel [100]$ and $y \parallel [010]$  are the in-plane axes. The value $2 \delta$ stands for the energy spacing between the
$E1$ and $HH1$ subbands in the absence of mixing; $\delta$ can be tuned from positive to negative value by varying the QW thickness $d$. In particular, $\delta=0$ for the critical thickness $d=d_c$~\cite{Gerchikov1989}. The parameter $A$ determines the velocity of Dirac fermions. In the $\bm k$$\cdot$$\bm p$ model, it is given by $A = (P / \sqrt{2}) \int f_1(z) f_3(z) dz + \delta A$, where $P$ is the Kane matrix element and $\delta A$ stands for the contributions from remote bands. Finally, $\gamma$ describes the coupling of $E1$ and $HH1$ states at $\bm{k}=0$ in zinc-blende-lattice QWs~\cite{Konig2008,Winkler20012}, $2|\gamma| \approx 15$~meV as it follows from Figs.~\ref{fig2} and \ref{fig3} neglecting the influence of additional interface states.  

Solution of the secular equation for the matrix Hamiltonian~(2) yields the energy spectrum
\begin{align}\label{dispersion}
E_{1,4} = \mp \sqrt{\delta^2+(A |\bm{k}| + \gamma)^2} \:, \nonumber \\
E_{2,3} = \mp \sqrt{\delta^2+(A |\bm{k}| - \gamma)^2} \:.
\end{align}
The corresponding wave functions are given by
\begin{equation}
\Psi_{1,4} = \frac12 \left( 
\begin{array}{c}
a_{1,4} \\
b_{1,4} \, {\rm e}^{-{\rm i} \varphi} \\
- a_{1,4} \, {\rm e}^{-{\rm i}\varphi} \\
b_{1,4}
\end{array}
\right) , \:
\Psi_{2,3} = \frac12 \left( 
\begin{array}{c}
a_{2,3} \\
b_{2,3} \, {\rm e}^{-{\rm i} \varphi} \\
a_{2,3} \, {\rm e}^{-{\rm i}\varphi} \\
- b_{2,3}
\end{array}
\right) ,
\end{equation}
where $a_{1,4}= \mp{\rm i} \sqrt{(E_{1,4}+\delta)/E_{1,4}} \, {\rm sign}(A k+\gamma)$, $a_{2,3} = \mp {\rm i} \sqrt{(E_{2,3}+\delta)/E_{2,3}} \, {\rm sign}(A k -\gamma)$, $b_{l}= \sqrt{(E_{l}-\delta)/E_{l}}$, $l$ is the branch index, $k=|\bm{k}|$, and $\varphi$ is the polar angle of the wave vector $\bm{k}$, $\mathrm e^{\mathrm i \varphi} = k_+/k$.
The electron dispersion \eqref{dispersion} is shown in Fig.~\ref{fig1}. In the structure of the critical thickness the spectrum consists of two Dirac cones shifted vertically with respect to each other by $2|\gamma|$. At $d \neq d_c$, the gap of $2|\delta|$ opens in the spectrum at the wave vector $k_0=|\gamma/A|$.

The large anticrossing of the $E1$ and $HH1$ subbands at $\bm k = 0$ revealed in the atomistic calculations indicates that the subband mixing mainly originates from the interface inversion asymmetry related to the anisotropy of chemical bonds.   
Since the $E1$ subband is formed from the electron and light-hole basis functions and the $HH1$ subband is of the heavy-hole type, see Eq.~\eqref{Bloch}, they are efficiently coupled due to heavy-hole--light-hole mixing at the interfaces, the effect known for zinc-blende-lattice QW structures~\cite{Ivchenko1996,Krebs1998,Magri2000,Toropov2000,Durnev2014}. 
This short-range mixing can be modeled by introducing interface terms to the effective Luttinger Hamiltonian which take into account low spatial symmetry of individual interfaces. The additional terms related to for the left ($L$) and right ($R$) interfaces have the form $V_{L,R} = \pm \hbar^2 t_{l\mbox{-}h} /(\sqrt{3} a_0 m_0)\{J_x J_y\}_s \delta(z - z_{L,R})$, where $t_{l\mbox{-}h}$ is the (real) mixing constant, $a_0$ is the lattice constant, $m_0$ is the free electron mass, $\{J_x J_y\}_s = (J_x J_y + J_y J_x)/2$, $J_{x}$ and $J_y$ are the matrices of the angular momentum $3/2$, and $z_{L,R}$ are the interface positions. The terms leads to the coupling of the $E1$ and $HH1$ subband with the parameter
\begin{equation}\label{gamma}
\gamma = \frac{\hbar^2 \, t_{l\mbox{-}h}}{2 a_0 m_0} \left[ f_3(z_R) f_4(z_R) - f_3(z_L) f_4(z_L) \right] \:.
\end{equation}
Since the envelope functions $f_3(z)$ and $f_4(z)$ have opposite parities, the parameter $\gamma$ is non-zero. Comparing the results of atomistic calculation with the $\bm k$$\cdot$$\bm p$ theory we obtain $t_{l\mbox{-}h} \approx 1.5$. Extrapolation to HgTe/Cd$_{0.7}$Hg$_{0.3}$Te QWs, the structures commonly used in experiment~\cite{Konig2007,Buttner2011,Olbrich2013}, gives $t_{l\mbox{-}h} \approx 1.1$ and $2|\gamma| = 10$~meV. 

The splitting of Dirac states may affect many phenomena, including weak localization, Shubnikov-de Haas oscillations and quantum Hall effect, spectra of THz radiation absorption, spin-flip Raman scattering, etc. The basic characteristic of electron systems is the single-particle density of states 
\begin{equation}
\rho(E) = \sum_{l , \bm{k}} \delta [E-E_l(\bm{k})] \:.
\end{equation}
For the QW system under study, it has the form
\begin{equation}
\rho(E) = \frac{|E|}{\pi A^2} \times
\left\{ 
\begin{array}{l}
0 , \; \mbox{if}~|E| \leq |\delta|, \\
|\gamma|/\sqrt{E^2 - \delta^2} , \;\mbox{if}~ |\delta| < |E| \leq \sqrt{\gamma^2 + \delta^2}, \\
1, \;\mbox{if}~ |E| > \sqrt{\gamma^2 + \delta^2}.
\end{array}
\right. \nonumber
\end{equation}

\begin{figure}[t]
  \includegraphics[width=0.75\columnwidth]{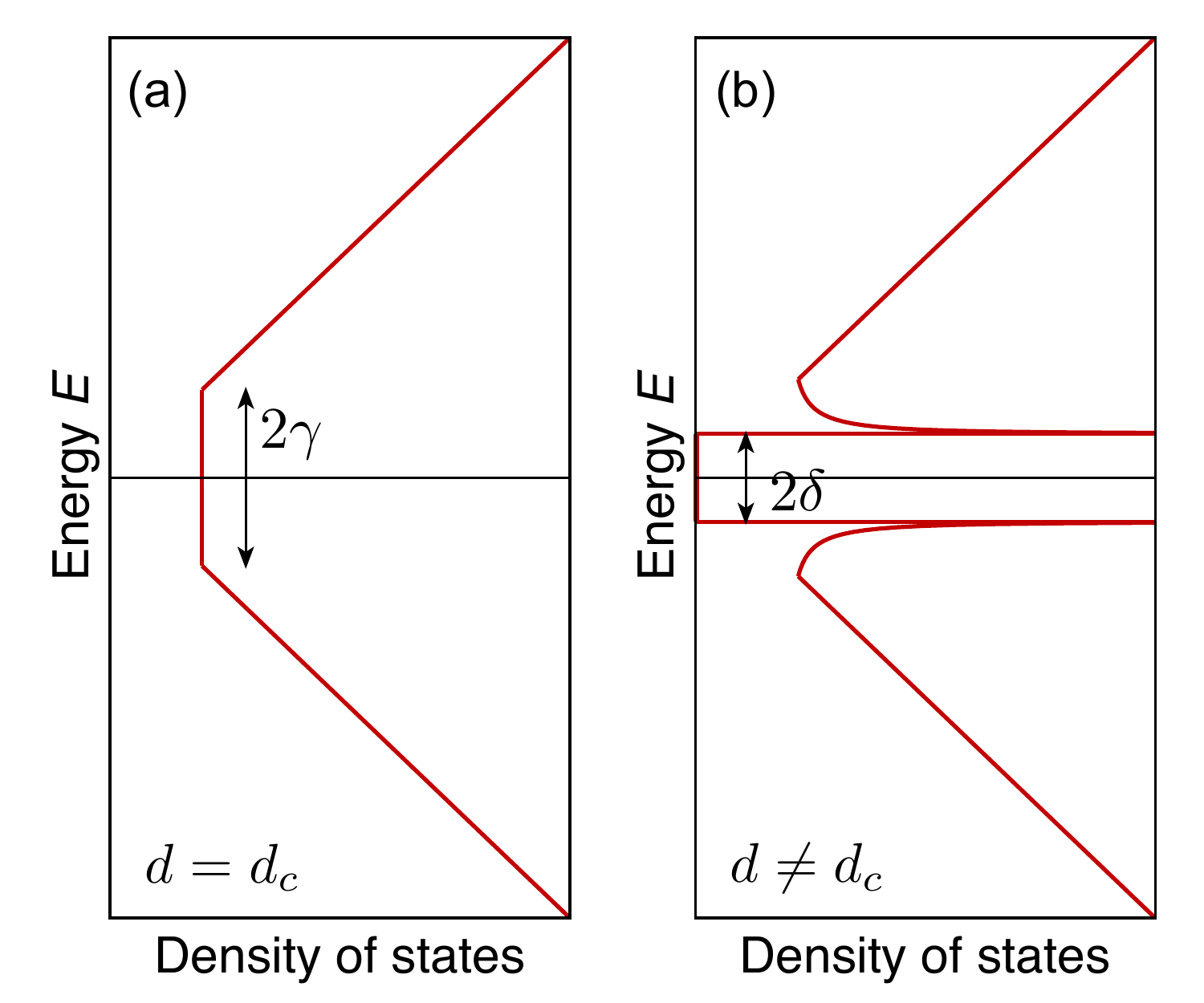}
\caption{Sketch of the density of states in HgTe/CdTe QWs of the (a) critical and (b) close-to-critical thinknesses.}
\label{fig_dos}
\end{figure}

Figure~\ref{fig_dos} illustrates the density of states vs. electron energy for QWs of the critical and close-to-critical thicknesses. The gapless structure behaves as a two-dimensional semimetal with the density of states remaining finite for the whole energy range, Fig.~\ref{fig_dos}a.  The density of states linearly scales with the energy $\rho(E)=|E|/(\pi A^2)$ at $|E|>|\gamma|$ and is energy independent, $\rho(E)=|\gamma|/(\pi A^2)$, at $|E|<|\gamma|$. In the latter region, electrons and holes coexist and the conductivity is bipolar. In the QWs of close-to-critical thicknesses, Fig.~\ref{fig_dos}b, the density of states has a gap of $2|\delta|$ and van Hove singularities at $E = \pm \delta$. 

The splitting of Dirac cones leads to a beating pattern in the Shubnikov-de Haas oscillations as well as splitting of the absorbance {peak in experiments on cyclotron resonance. The peak positions are determined by the effective cyclotron masses $m_l = \hbar^2 k /(d E_l /dk)$. In $n$-doped structure with the Fermi energy $E_F$, one can expect the splitting of resonant line into two components, their positions at $E_F \gg |\gamma|$ being determined by the effective masses $m_{3,4}=(\hbar/A)^2(E_F \pm \gamma)$.

The energy spectrum can be also studied by means of optical spectroscopy, particularly by measuring the spectral dependence of the QW absorbance. For direct optical transitions between occupied and empty states in the geometry of normal incidence of the radiation, the QW absorbance is given by 
\begin{equation}
\eta = \frac{4 \pi^2 \alpha \,\hbar}{\omega n_{\omega}} \sum_{l, m,\bm{k}} |\bm{e} \cdot \bm{v}_{ml}|^2 \, \delta(E_{m} - E_{l} -\hbar\omega) \:,
\end{equation}
where $\alpha$ is the fine-structure constant, $\omega$ is the light frequency, $n_{\omega}$ is the refractive index of the
medium, $\bm{e}$ is the (complex) unit vector of the light polarization, and $\bm{v}_{ml}=\hbar^{-1}(\partial H /\partial \bm{k})_{ml}$ is the matrix element of the velocity operator. We consider undoped QW structures where the optical transitions can occur between the occupied branches $1$ and $2$ and the empty branches $3$ and $4$ of the energy spectrum, see insets in Figs.~\ref{fig_absorbance}a and~\ref{fig_absorbance}b. 
\begin{figure}[t]
  \includegraphics[width=0.8\columnwidth]{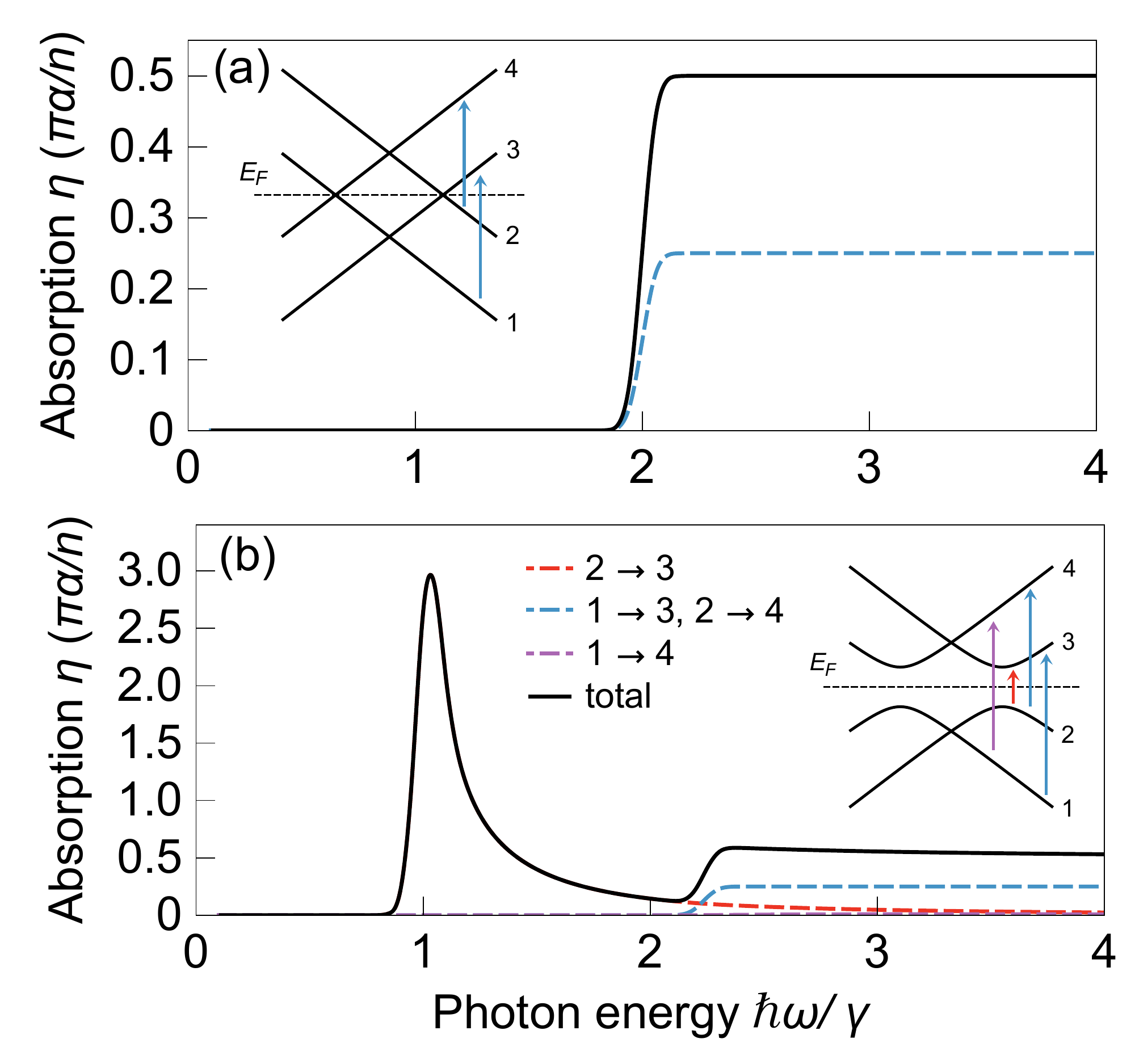}
\caption{Photon energy dependence of the QW absorbance for the structures of (a)  the critical thickness, $\delta=0$, and (b) the close-to-critical thickness, $\delta = 0.5 \gamma$. Dashed curves show partial contributions to the absorbance. Inset sketches the energy spectrum and allowed optical transitions. The steep spectral edges are smoothed by convoluting the spectra with the Gauss function with the standard deviation of $0.05 \gamma$.}
\label{fig_absorbance}
\end{figure}
For normally-incident radiation, the QW absorbance is independent of the light polarization and has the form 
\begin{equation}
\eta = \frac{\pi \alpha A^2}{\hbar \omega n_{\omega}} \sum_{\substack{l=1,2 \\ m=3,4} } \sum_{k_t} \frac{k_t f_{ml} (k_t)}{|dE_{m}/dk - dE_{l}/dk|_{k=k_t}} \:,
\end{equation}
where $k_t$ is the absolute value of the wave vector at which the optical transitions between the branches $l$ and $m$ occur, it is found from the equation $E_{m} (k_t)-E_{l}(k_t)=\hbar\omega$, and $f_{ml}(k)$ are the dimensionless functions defined by 
$f_{ml}(k)=(\hbar/A)^2|\bm{e} \cdot \bm{v}_{ml}|^2$. Straightforward calculations yield
\begin{eqnarray}\label{f_coefficients}
&&f_{41} = \frac{\delta^2}{E_{4}^2} \:, \;\;\; f_{32} =  \frac{\delta^2}{E_{3}^2} \:, \\
&&f_{31} (k) = f_{42} (k) =  \frac12 \left( 1 - \frac{A^2 k^2 - \gamma^2 + \delta^2}{E_1 E_3} \right)\:. \nonumber
\end{eqnarray}
The functions $f_{31}(k)$ and $f_{42}(k)$ 
change from $\delta^2/(\delta^2 + \gamma^2)$ at $k=0$ to close to unity at $k > |\gamma/A|$.

Figures~\ref{fig_absorbance}a and~\ref{fig_absorbance}b demonstrate the photon energy dependence of the QW absorbance for structures of the critical and close-to-critical thicknesses. In the structure with two split Dirac cones, Fig.~\ref{fig_absorbance}a, the absorption spectrum has a step-like shape with the edge at $\hbar\omega = 2|\gamma|$ although there is no band gap in the energy spectrum. Such a behavior is dictated by the selection rules and energy conservation law: Direct optical transitions in QWs of the critical thickness are allowed only between the branches $1 \rightarrow 3$ and $2 \rightarrow 4$ and these transitions can occur at $\hbar\omega \geq 2|\gamma|$.  In QWs of non-critical thickness, Fig.~\ref{fig_absorbance}b, the branches $2$ and $3$ anticross at finite wave vector, and the direct optical transitions between them get allowed. It leads to the emergence of an additional sharp band in the absorption spectrum at $\hbar\omega = 2|\delta|$. The spectral shape of this absorption band is determined by the van Hove singularities in the density of states.

To summarize, we have described the splitting of Dirac states in HgTe/CdTe quantum wells of critical and close-to-critical thicknesses.
In structures of the critical thickness, the splitting between the Dirac cones reaches the value of $15$~meV and is dominated by symmetry-enforced light-hole--heavy-hole mixing at the quantum well interfaces. These structures behave as a two-dimensional semimetal with non-vanishing density of states in the whole energy range. In quantum wells of close-to-critical thicknesses, a gap opens at a finite in-plane wave vector, which leads to the emergence of extremum circles in the electron dispersion and corresponding van Hove singularities in the density of states. We have also discussed how the peculiarities of the energy spectrum may be revealed in optical and transport measurements.

Financial support by the RFBR, RF President Grants MD-3098.2014.2 and NSh-1085.2014.2, EU project SPANGL4Q, and the ``Dynasty'' Foundation is gratefully acknowledged. Work at University of Colorado, Boulder by A.\,Z.  was supported by the U.S. Department of Energy, Office of Science, Basic Energy Science, Materials Sciences and Engineering Division under Contract No. DE-FG02-13ER46959 to University of Colorado.

\newpage

\section*{SUPPLEMENTARY MATERIAL}

%
%
%
%

The tight-binding parameters used for the calculation of the HgTe/CdTe quantum well band structure are presented in Table~\ref{tbl:1}.
The parameters are fitted to describe the band structures (energy gaps and effective masses) of HgTe and CdTe bulk crystals.
The energy gaps were calculated in the WIEN2k package~\cite{WIEN2k} using modified Becke-Johnson (mBJ) exchange-correlation potential~\cite{MBJ} which is known to give a good agreement with experimental data~\cite{MBJTI}. The effective masses are determined from the band structure parameters given in Ref.~\cite{Novik05}.
The energy gaps and effective masses obtained in the tight-binding model with the parameters listed in Table~\ref{tbl:1} and corresponding reference values are compared in Table~\ref{tbl:2}. The difference in lattice constants is neglected, the lattice constant $a=6.453$~\AA{} is used.

\begin{table}[h]
    \begin{tabular}{|c|cc|}
\hline
                 &       HgTe &       CdTe \\
\hline
$         E_{sa}$ & $   -8.3491$ & $   -9.4200$ \\
$         E_{sc}$ & $   -3.8332$ & $   -1.2731$ \\
$         E_{pa}$ & $    2.1509$ & $    0.5800$ \\
$         E_{pc}$ & $    1.9005$ & $    2.9190$ \\
$       ss\sigma$ & $   -1.3560$ & $   -1.2125$ \\
$   s_ap_c\sigma$ & $    2.6871$ & $    2.2262$ \\
$   s_cp_a\sigma$ & $    1.2910$ & $    2.4611$ \\
$       pp\sigma$ & $    2.9785$ & $    2.9661$ \\
$          pp\pi$ & $   -0.6403$ & $   -0.6213$ \\
$       \Delta_a/3$ & $    0.2907$ & $    0.3303$ \\
$       \Delta_c/3$ & $    0.1984$ & $    0.0595$ \\
\hline
\end{tabular}
\caption{Tight-binding parameters (in eV) of HgTe and CdTe bulk crystals.}
\label{tbl:1}
\end{table}

\begin{table}[h]
\begin{tabular}{|c|cc|cc|}
\hline
                 &       \multicolumn{2}{c|}{HgTe}     & \multicolumn{2}{c|}{CdTe}  \\
\hline
                 &       TB  & Reference       &  TB & Reference \\
\hline
$E(\Gamma_{6}^v) $ &$-11.9665$ &$-11.97$\footnotemark[1] & $-11.6801$& $-11.68$\footnotemark[1] \\
$E(\Gamma_{7}^v) $ &$ -0.7275$ &$-0.728$\footnotemark[1] & $-1.3523$ & $-1.347$\footnotemark[1] \\
$E(\Gamma_{8}^v) $ &$  0.0000$ &$ 0.000$                 & $-0.5756$ & $-0.570$\footnotemark[2] \\
$E(\Gamma_{6}^c) $ &$ -0.2158$ &$-0.217$\footnotemark[1] & $ 0.9870$ & $ 0.987$\footnotemark[1] \\
$E(\Gamma_{7}^c) $ &$  3.8007$ &$ 3.510$\footnotemark[1] & $ 4.0717$ & $ 3.500$\footnotemark[1] \\
$E(\Gamma_{8}^c) $ &$  4.5406$ &$ 4.683$\footnotemark[1] & $ 4.4644$ & $ 4.426$\footnotemark[1] \\
\hline
$m_{lh}^{001}$ &$ 0.0268 $& $0.027$\footnotemark[2] & $-0.1094$& $-0.115$\footnotemark[2]\\
$m_{lh}^{111}$ &$  0.0281$& $0.029$\footnotemark[2] &$-0.0969 $& $-0.107$\footnotemark[2]\\
$m_{hh}^{001}$ &$-0.3155$ & $-0.32$\footnotemark[2]  &$-0.4022$ &$ -0.493$\footnotemark[2]\\
$m_{hh}^{111}$ &$-0.6703$ & $-0.67$\footnotemark[2]  &$-0.7628$ &$ -0.709$\footnotemark[2] \\
$m_{e}^{001}$  &$-0.0319$ & $-0.031$\footnotemark[2] &$ 0.0960$ &$ 0.090$\footnotemark[2]\\
$m_{e}^{111}$  &$-0.0319$ & $-0.031$\footnotemark[2] &$ 0.0960$ &$ 0.090$\footnotemark[2]\\\hline
\end{tabular}
\footnotetext[1]{Values obtained in WIEN2k~\cite{WIEN2k} with mBJ functional~\cite{MBJ}}
\footnotetext[2]{Values deduced from Ref.~\cite{Novik05}}
\caption{Reference band structure parameters and the parameters obtained in the tight-binding model using parameters of Table~\ref{tbl:1}.
}\label{tbl:2}
\end{table}

We use the $sp^3$ tight-binding model where all parameters may be unambiguously extracted from the electron energies in the high symmetry points in the Brillouin zone. We are inspired by the parametrization procedure described in Ref.~\cite{Vogl} for A$_3$B$_5$ materials. For A$_2$B$_6$ materials, where the valence band is significantly mixed with the lower $d$ band, we modify the procedure: We use the masses in the valence band instead of the energies $E_v(X)$ as reference parameters.

\end{document}